\documentclass{jaa}
\usepackage{natbib}
\usepackage{aas_macros}
\usepackage{url}
\usepackage{graphicx}
\usepackage{hyperref}
\begin{document}\sloppy

\title{Real-time RFI Filtering for uGMRT: Overview of the Released System and Relevance to the SKA}

\author{Kaushal D. Buch\textsuperscript{1,*}, Ruta Kale\textsuperscript{1}, Mekhala Muley\textsuperscript{1}, Sanjay Kudale\textsuperscript{1}, Ajithkumar B.\textsuperscript{1} }
\affilOne{\textsuperscript{1}Digital Backend Group, Giant Metrewave Radio Telescope, NCRA-TIFR, Pune 410504, India.\\}

\twocolumn[{

\maketitle

\corres{kdbuch@gmrt.ncra.tifr.res.in}

\msinfo{19 March 2022}{12 December 2022}

\begin{abstract}
Radio Frequency Interference (RFI) of impulsive nature is created by sources like sparking on high-power transmission lines due to gap or corona discharge and automobile sparking, and it affects the entire observing frequency bands of low-frequency radio telescopes. Such RFI is a significant problem at the Upgraded Giant Metrewave Radio Telescope (uGMRT). A real-time RFI filtering scheme has been developed and implemented to mitigate the effect on astronomical observations.
The scheme works in real-time on pre-correlation data from each antenna and allows the detection of RFI based on median absolute deviation statistics. 
The samples are identified as RFI based on user-defined thresholds and are replaced by digital noise, a constant or zeros. We review the testing and implementation of this system at the uGMRT. We illustrate the effectiveness of the filtering for continuum, spectral line and time-domain data. The real-time filter is released for regular observations in the bands falling in 250 - 1450 MHz, and recent observing cycles show growing usage. Further, we explain the relevance of the released system to the Square Kilometer Array (SKA) receiver chain and possible ways of implementation to meet the computational requirements.

\end{abstract}


}]

\doinum{12.3456/s78910-011-012-3}
\artcitid{\#\#\#\#}
\volnum{000}
\year{0000}
\pgrange{1--}
\setcounter{page}{1}
\lp{1}

\section{Introduction}
Radio Frequency Interference (RFI) is a term used for man-made signals in radio bands that are inevitably picked up by radio telescopes trying to study radio signals from celestial sources. RFI limits our ability to measure cosmic signals, and thus its mitigation is necessary. While we are building radio telescopes with unprecedented sensitivities, the RFI is also increasing due to the increasing electronic communication from terrestrial and extra-terrestrial sources \citep[e. g.][]{baan2010rfi, ford2014rfi,thompson2017radio}. Thus, RFI mitigation is still needed despite locating 
the radio telescopes in remote locations with scarce human populations. Moreover, the telescope receiver systems and other on-site devices can also generate RFI that needs to be measured and excised.

RFI is broadly classified as ``broadband" and ``narrowband" categories based on the spectral properties \citep[e. g.][]{ford2014rfi}.  The effect of broadband RFI is spread across the entire observing frequency band, while that of narrowband RFI is localized to narrow ranges of frequencies within the observing band. Typical sources of broadband RFI include powerline sparking due to gap or corona discharge and automobile sparking. The sources of narrowband RFI include broadcast transmitters, navigational satellites, and wireless communication from mobile phones and short-range transmitters.

RFI mitigation has been explored in many ways and implemented based on the requirement of the observatories. The most widely used techniques operate on the data after it has been recorded; we refer to these as 
offline techniques. PIEFLAG is a python-based tool to identify RFI from a dataset using amplitude-based and RMS-based methods \citep{2006PASA...23...64M}. Another tool which is widely used on Murchison Widefield Array data and LOFAR is AO-Flagger, which is effective on narrowband RFI \citep{2015PASA...32....8O}. More recently, methods based on Deep Fully Convolutional Neural Networks (DFCN) \citep{2019MNRAS.488.2605K}  and Convolutional Neural Networks (CNN) \citep{2022MNRAS.512.2025S} have also been developed for improving the speed and efficacy of RFI identification and flagging. These have been mainly used and tested on telescopes mainly affected by narrowband RFI. For mitigating broadband RFI, methods that operate in real-time, as the data are recorded, are needed.

The Upgraded Giant Metrewave Radio Telescope (uGMRT) \citep{2017CSci..113..707G} is a pathfinder of the Square Kilometer Array (SKA) and is located at Khodad, about 90 km. North of Pune in India. 
It consists of 30 dish antennas distributed roughly in a ``Y"-shaped array \citep{swarup1991giant} with maximum and minimum baseline lengths of $25$ km and $100$ m, respectively. The ``upgrade" of the GMRT involved the installation of new receivers and a wideband correlator to allow data recording with broad instantaneous bandwidths \citep{2017CSci..113..707G}.
The uGMRT is a system offering broad instantaneous bandwidths of up to 400 MHz.
Furthermore, it operates in four frequency bands, namely, 120 - 240 MHz (Band 2), 250 - 500 MHz (Band 3), 550 - 850 MHz (Band 4) and 1000 - 1450 MHz (Band 5). 

This paper focuses on the broadband RFI at the uGMRT, the real-time mitigation (filtering) system implementation and the effect of filtering on astronomical observations. Further, we motivate ideas for the SKA, which proposes a threshold-based RFI detection and filtering system in the signal processing chain. The review is organized as follows. We provide an overview of the broadband RFI at the uGMRT in Sec.~\ref{sec:ugmrtrfi}. The real-time RFI filtering technique is described in Sec.~\ref{sec:techreview}. A review of the tests carried out is presented in Sec.~\ref{sec:testscheme}, followed by the description of the impact of the filter on astronomical data in Sec.~\ref{sec:astro}. 
The current usage statistics for the RFI filtering are reported in Sec.~\ref{sec:usage}. The relevance of real-time filtering is discussed in Sec.~\ref{sec:skarel}. Conclusions are presented in Sec.~\ref{sec:conclusions}.

\section{uGMRT and RFI}\label{sec:ugmrtrfi}

The GMRT Wideband Backend (GWB) \citep{reddy2017wideband} consists of 16 ROACH-1 boards, which carry out real-time RFI filtering and packetize the data into 10 Gigabit Ethernet(GbE) packets for further processing using a 16-node GPU cluster. GWB operates in the correlation and beamforming modes. Further, the beamforming can be carried out using Incoherent Array and Phased Array modes \citep{reddy2017wideband}.

Powerline RFI at the GMRT is caused due to sparking on high-power transmission lines and distribution equipment. Many domestic power lines for residential and farming requirements are located near the array. Apart from this, 11 kV, 33 kV, HVAC, and 500 kV DC lines cross the outermost arm antennas \citep{raybole2010external}. Several potential sources of RFI are located around the central square and near the arm antennas of the GMRT array \citep{swarup2008power}.

\begin{figure*}
    \centering
    \includegraphics[, trim = 4cm 0.6cm 4cm 1cm, clip, width=17cm]{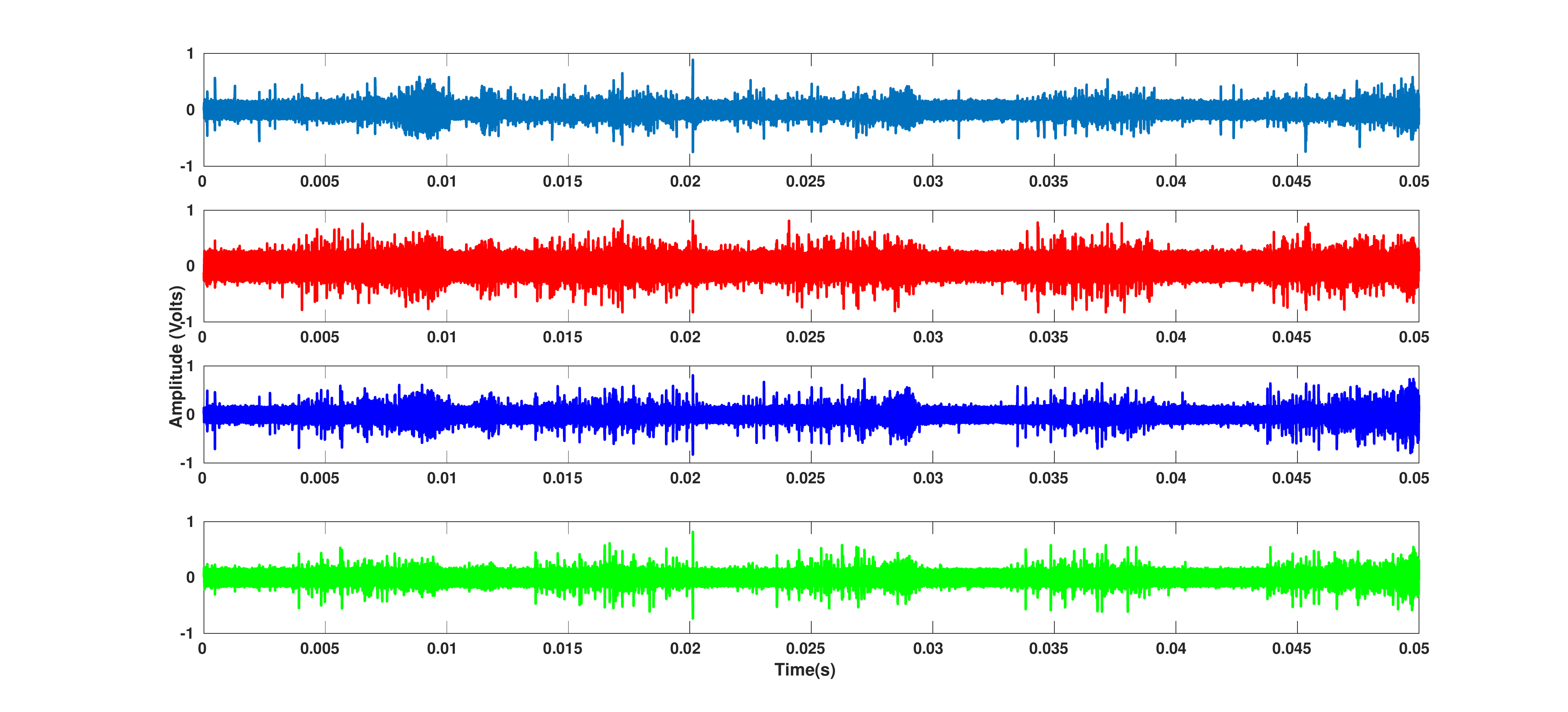}
    \caption{Oscilloscope output at five ns sampling from C05 and C06 antennas of the GMRT at band-4 is shown. This 50 ms period trace of amplitude versus time was recorded at the input of the GWB for two polarizations of C05 and C06 antennas ($\sim$100 m baseline). The top two subplots show C05 (dark blue color) and C06 (red color) for Pol. 1, respectively, and the bottom two subplots show the same antennas, i.e. C05 (blue color) and C06 (green color) for Pol. 2, respectively. The powerline RFI has a periodicity of 10ms and is correlated across the antennas and polarizations.}
    \label{corrrfi}
\end{figure*}

Powerline RFI generally manifests as a bunch of impulses in the time domain and has properties similar to gap discharge \citep{maruvada2000corona}. These bunches show a periodicity and usually repeat at multiples of powerline frequency (50 Hz in India)\citep{swarup2008power}. The typical periodicity is 10 ms. This RFI depends on observing frequency, and its strength is significant in bands 2, 3, and 4 of the uGMRT. The strength, duration, and number of impulses in the bunch depend on the moisture, wind, and faults. A typical time-series plot of a signal affected by powerline RFI is shown in Fig. \ref{corrrfi}. 

Powerline RFI is $10 - 20$ dB stronger than the system noise and leads to broadband RFI increase in the power level in the power spectrum of the astronomical signal. Impulses from powerline RFI are outliers and make the Gaussian distribution of the signal heavy-tailed. This type of RFI is spatially correlated and shows a strong signature on short baselines, as seen in Fig. \ref{corrrfi}.

\subsection{Need for Real-time Filtering}

Real-time RFI filtering at the highest time resolution helps prevent data corruption in downstream signal processing with minimal loss of astronomical data. The benefits of having a real-time filtering system in the digital backend are as follows:

\begin{enumerate}
    
\item The energy in impulsive time-domain RFI spreads across the observing band upon Fourier transformation, making it difficult to detect and mitigate in the post-processing operation.

\item Since the RFI is broadband, tuned or notch filters are not helpful for mitigation, even if used in the earlier stages of the receiver system.

\item Mitigation in the pre-correlation domain helps reduce the ill effects of correlated RFI.

\end{enumerate}

\section{Real-time RFI Filtering Technique}\label{sec:techreview}

The real-time RFI filtering technique developed and implemented \citep{buch2019real} at the uGMRT is based on detection and excision \citep{buch2019rfi} of interference. 
The RFI filtering operation is divided into three basic steps - Estimation, Detection, and Filtering. These operations are carried out on each antenna and polarization data. The robustness and performance of MAD-based (Median Absolute Deviation) estimators in detecting impulsive RFI are shown in \citet{buch2016towards}. Detailed architecture and implementation are provided in \citet{buch2019real}. The filter can also operate in detection-only mode, bypassing the filtering operation.

\begin{enumerate}

\item \textbf{Estimation:}
Median Absolute Deviation (MAD) is computed on Nyquist-sampled digitized time series ($\rm{MAD} = \rm{median} (\rm{abs}(X_i - \rm{median}(X)))$) on 16k data samples. In order to get an unbiased estimate, the minimum window size should be twice the expected duration of RFI burst \citep{buch2016real}.
 For longer bursts of RFI, the median of 16k MAD values are computed for getting Median-of-MAD
(MoM), which we refer to as 16k MoM. This method is a trade-off between estimating a long duration and hardware resource utilization.
Further, MoM is computed on every fourth input sample out of the samples arriving at 1.25 ns intervals to meet the real-time requirements. Thus, the estimation interval for 16k MoM is 1.34 s which can be calculated as $(16384 \rm{samples}*16384 \rm{MAD} \rm{values}*1.25*10^{-9} \rm{sampling period})*4 = 1.34)$s. 

\item \textbf{Detection:}
The upper and lower thresholds are computed from the MoM value as
$\rm{median} \pm n*1.4826*\rm{MoM}$. The factor $n$ is programmable.
The two detection techniques designed for the RFI filtering system are (a) Voltage detection and (b) Power detection. The voltage detection technique compares the values of the individual samples with the thresholds, whereas the power detection technique uses the squared and accumulated samples for comparison with the threshold. The threshold mentioned above is used for voltage detection, whereas an appropriate scaling is carried out to extend the threshold for power detection (based on the number of samples accumulated).

At this stage, the flagging statistics are recorded using an RFI counter that keeps a record of the number of detected
RFI samples and a total number of samples per antenna and
polarization. These counts would provide the fraction of flagged samples over user-defined time intervals during an observing run.

\item \textbf{Filtering:}
The samples detected as RFI can be replaced with digital noise, a constant value, or a threshold. The user can choose a replacement option. 
Samples from a standard normal distribution are generated on Field Programmable Gate Array (FPGA) \citep{buch2014variable} and are scaled to the estimated statistics of the input signal. 
\end{enumerate}
\subsection{Implementation}
The RFI filtering system is an FPGA-based implementation on a ROACH-1 board\footnote{\url{https://casper.ssl.berkeley.edu/wiki/ROACH}} from CASPER (Collaboration for Astronomy Signal Processing and Electronics Research) \citep{hickish2016decade}. 

\begin{figure*}
    \centering
    \includegraphics[trim = 0cm 2cm 0cm 3cm,clip,width=18cm]{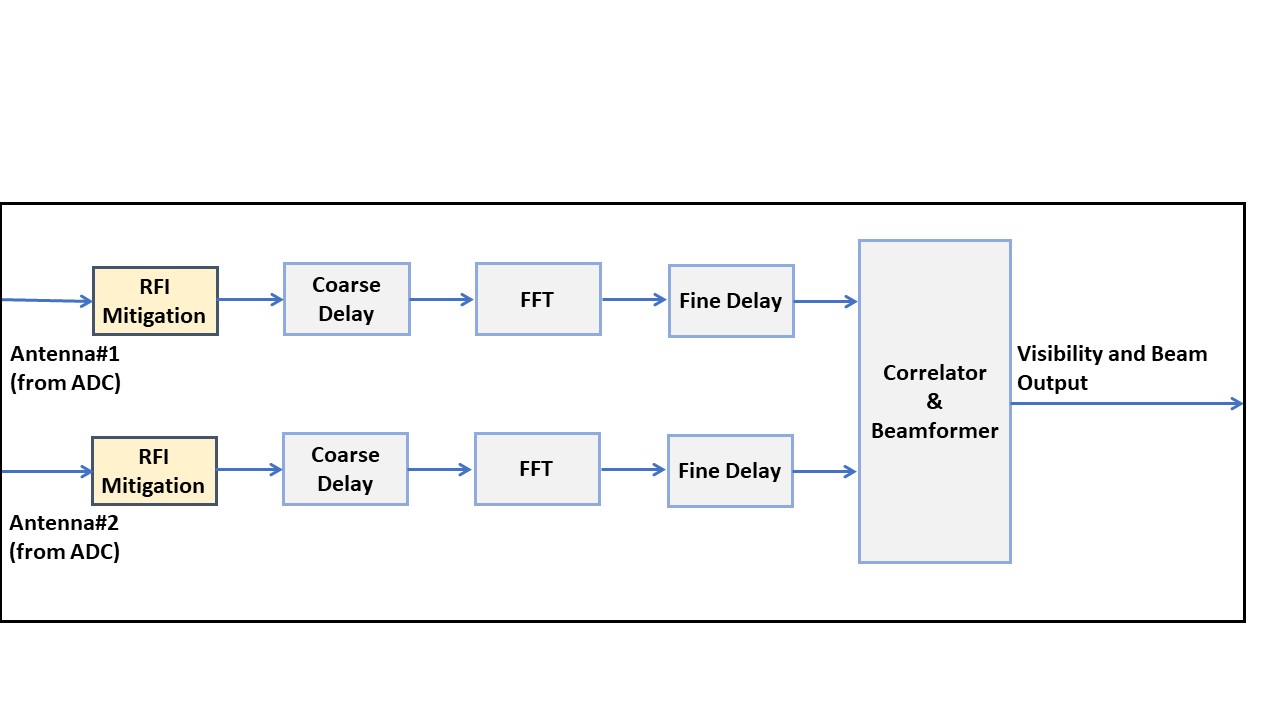}
    \caption{
    A simplified block diagram of a 2-element correlator/beamformer for GWB showing the location of the real-time RFI filtering system is presented.}
    \label{blkdia}
\end{figure*}

The system is implemented on the ROACH-1 board at the start of the signal processing chain. A simplified block diagram of the chain showing the location of the filtering system is shown in Fig. \ref{blkdia}. Each ROACH-1 board processes four inputs, digitized to an 8-bit precision at 800 MHz sampling frequency. Signals from four different antennas of a particular polarization are fed to a single ROACH-1 board. A total of sixteen ROACH boards cover the thirty dual-polarized GMRT antenna inputs. The RFI filtering system is programmed and configured through a control computer via a 1 Gigabit Ethernet (GbE) interface. The parameters of the filtering system on each board can be configured separately through a software interface. RFI counter readout is also through the 1 GbE interface.

An earlier design version is available in the CASPER open-source library \footnote{\url{https://casper.astro.berkeley.edu/wiki/Projects}}.
Hardware optimization and real-time operation are achieved using the histogram method \citep{buch2019real} for median computation. Other trade-offs and optimization are described in \citet{buch2019real}.

\section{Review of testing schemes}\label{sec:testscheme}

The real-time RFI filtering system tests are carried out at two levels - Engineering and Astronomical Tests. A simultaneous recording of unfiltered and filtered data is carried out for a fair comparison. The inputs from 30 antennas are split to get 15 unfiltered and 15 copies that undergo filtering. We call this the 1:2 digital copy mode \cite{buch2022performance}. The digital backend system is appropriately programmed to treat copies from the corresponding unfiltered counterparts. A detailed description of the 1:2 digital copy mode, including the data recording and acquisition process for correlator and beamformer outputs, is provided in \citep{buch2018implementing, buch2022performance}. Similarly, for comparing multiple filter configurations simultaneously, 28 antenna inputs are split into seven unfiltered copies and seven each with different filter configurations (different threshold or replacement options) to get 1:4 digital copy mode \citep{buch2022performance}.

The digital copy modes are also used to observe the effect of threshold and replacement options on the Signal-to-Noise ratio (SNR) and compare the data between the beam-former and correlator modes \citep{buch2018implementing}. These tests help fine-tune the threshold and replacement options for optimizing the filter performance.

\subsection{Engineering Tests}
The tests are carried out to compare the raw voltage, spectrogram, beamformer, and correlator (visibility) outputs. 

We provided controlled inputs to the filtering system for characterization during the initial testing phase. We could change the duration and strength of RFI and characterize the filtering in the 1:2 digital copy mode. Examples of results are shown in \citep{buch2018implementing}. Such controlled testing allows for setting an appropriate input power level and setting a bound on the filter's performance.

More detailed tests are carried out on calibrator radio sources to analyze the effect on SNR, suppression of 50 Hz powerline signature, cross-correlation function, and closure phase. These tests were carried out for a wide range of receiver parameters like bandwidth, observing band, number of spectral channels and integration time.

\subsubsection{Effect on SNR}

For analyzing the effect on SNR, we carry out a simultaneous comparison between unfiltered and filtered spectrograms (magnitude spectrum over a specific time interval) for a GMRT antenna. We study the temporal behaviour of a single frequency channel. A quantitative comparison between the theoretical SNR given by the radiometer equation is made, particularly for instances affected by RFI. The ratio of SNR before filtering to the one after filtering measures the improvement achieved through the filtering system. The improvement depends on the intensity of RFI and the filtering efficiency. Through several test observations carried out in the different uGMRT observing bands, we have observed an improvement in the SNR ranging from 0.5 dB to 10 dB.

\subsubsection{Suppression of 50 Hz and harmonics}

The presence of powerline RFI manifests as 50 Hz and harmonics in the Fourier transform of a single frequency channel computed as a function of time. The magnitude transforms of unfiltered and filtered data are compared to determine the amount of suppression in the fundamental (50 Hz) and harmonics. The ratio of power in the 50 Hz for unfiltered data to the filtered data provides the quantitative measure of the RFI suppression. We have observed a typical suppression of 6-10 dB \citep{buch2022performance}.

\begin{figure*}
    \centering
    \includegraphics[trim=2cm 0.5cm 1cm 1cm,clip,width=18cm]{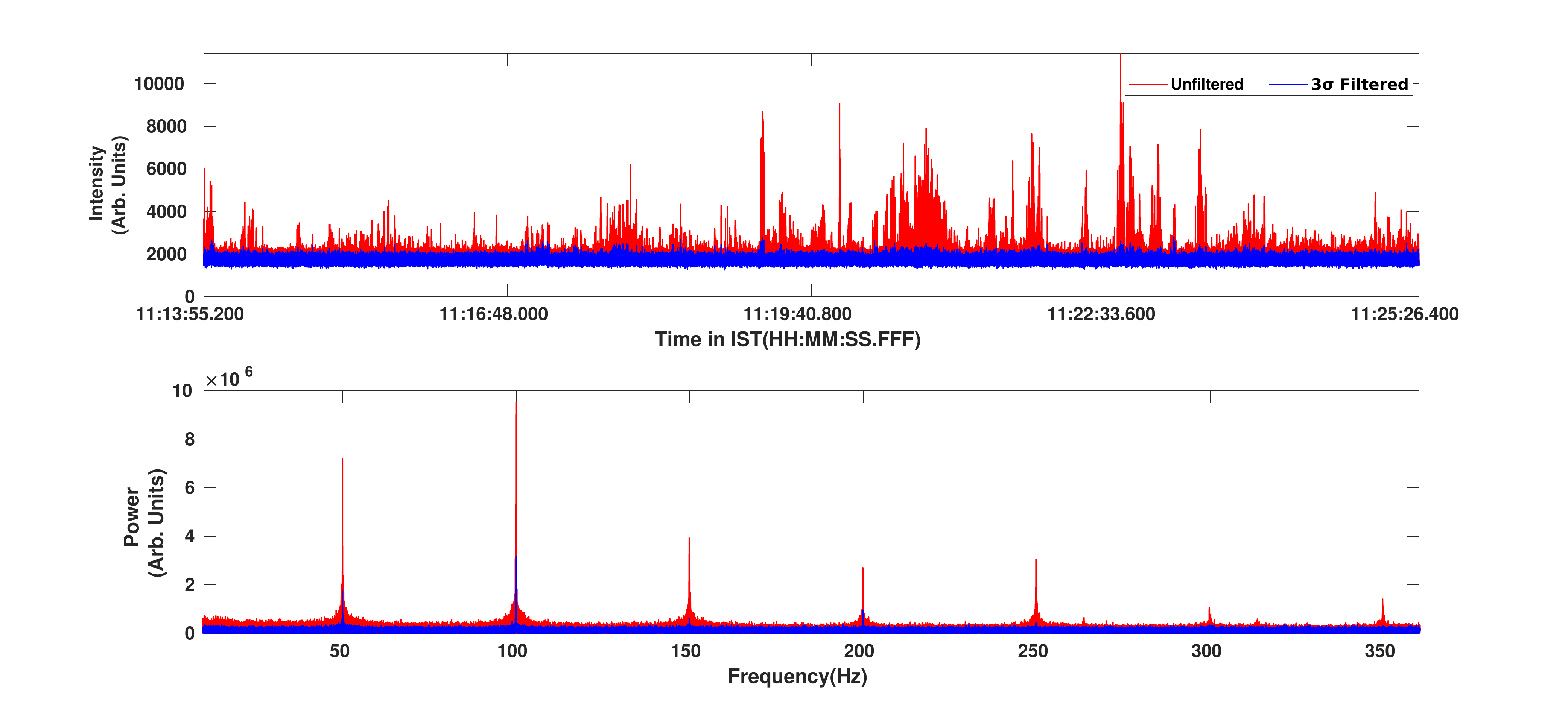}
    \caption{Beam output and its FFT for a single frequency channel corresponding to 431 MHz. The data is from the C09 antenna of GMRT for a 20 min observation of 3C147 in band-3, 200 MHz bandwidth, 2048 spectral channels, and 1.3ms integration time.}
    \label{beam_fft}
\end{figure*}

A comparison of power from a GMRT antenna for a spectral channel while observing a calibrator source (unresolved) in uGMRT Band-3 is shown in Fig. \ref{beam_fft}. The observation was carried out in 1:2 digital copy mode. The filtering of RFI is seen along with suppression in the 50 Hz and harmonics.

\subsection{Effect on the Cross-correlation Function (CCF)}

The uncertainty in the CCF measurement increases in the presence of RFI. We simultaneously compare the CCF across different baselines for time-averaged visibilities \citep{buch2022performance} to observe the uncertainty. Increased power fluctuations due to RFI lead to increased uncertainty in the measurement. Such fluctuations are significant at baselines less than 1 km. and lead to correlated RFI even when the antennas are pointing off-source \citep{buch2018implementing}. The amount of reduced uncertainty in the filtered data indicates an improvement in the visibility data. 

\begin{figure*}
    \centering
    \includegraphics[trim=3cm 0.5cm 1cm 0.5cm,clip,width=18cm]{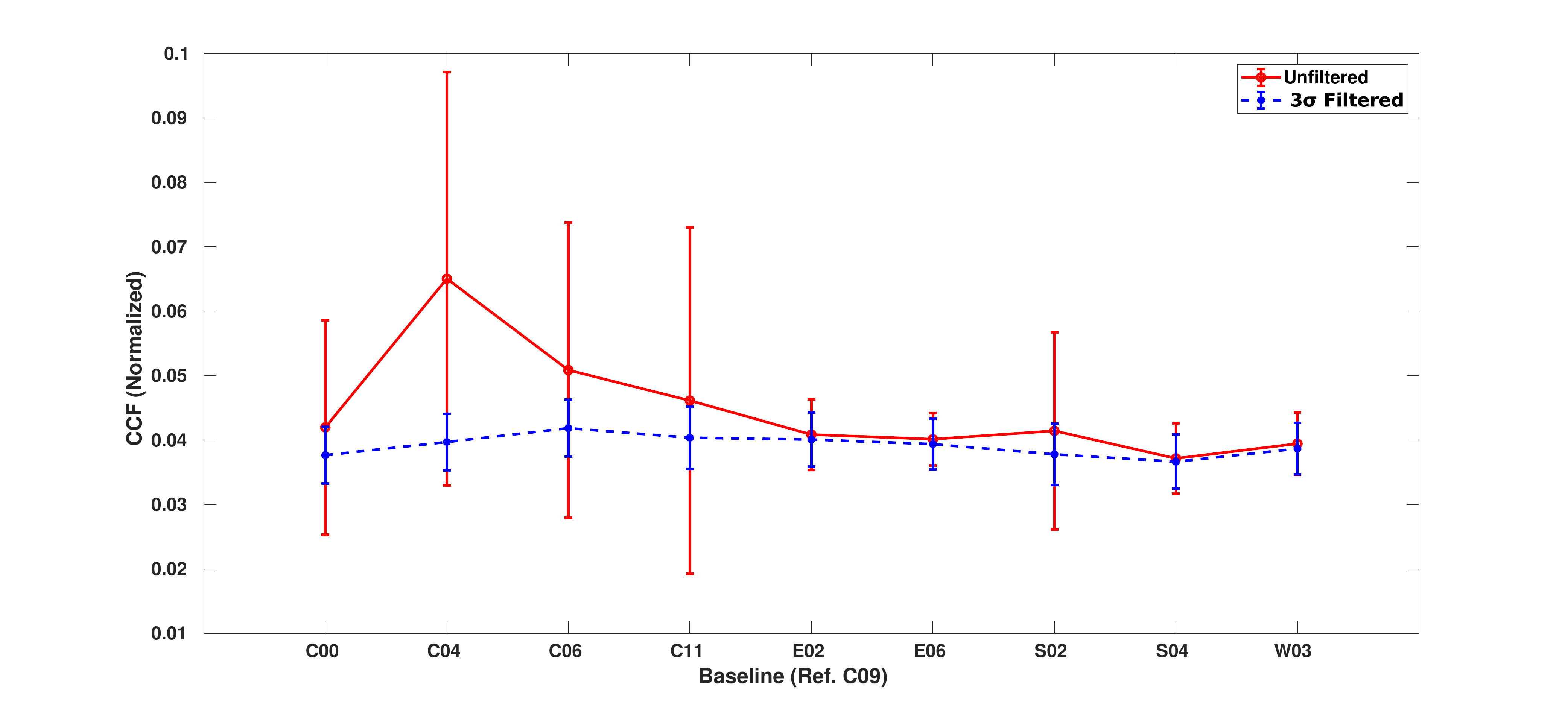}
    \caption{CCF of a single frequency channel corresponding to 651 MHz is shown for baselines involving the labelled antenna with C09. The data is from a 20 min observation of 3C468.1 in band-4, ten antennas, 100 MHz bandwidth, 2048 spectral channels, and 0.671s integration time. The approximate baseline lengths in metres with respect to C09 antenna are C00 (654m), C04 (678m), C06 (110m), C11 (674m), E02 (3009m), E06 (13076m), S02 (4383m), S04 (9378m), and S06 (13996m).}
    \label{cp}
\end{figure*}

A comparison of CCF for a calibrator source (unresolved) observed in uGMRT Band-4 is shown in Fig. \ref{cp}. The observation was carried out in 1:2 digital copy mode. The increased CCF on shorter baselines is due to broadband RFI, which is reduced after filtering.

\subsection{Effect on Closure Phase}
This test helps understand the effect of broadband RFI on the closure phase relations. Since the powerline RFI is prominent on shorter baselines, we choose three antennas and corresponding filtered copies. A comparison is made between the closure phase of the baselines computed from the cross-correlation data for unfiltered and filtered data. The closure phase departs from zero for data corrupted by RFI. This measurement is independent of any instrumental effects and can help study the effect of RFI on astronomical data. We observed a typical variation of 70 to 100 degrees in phase in the event of RFI, which reduces to about 15 degrees after filtering \citep{buch2022performance}.

\begin{figure*}
    \centering
    \includegraphics[trim=3cm 1cm 1cm 0.5cm,clip,width=18cm]{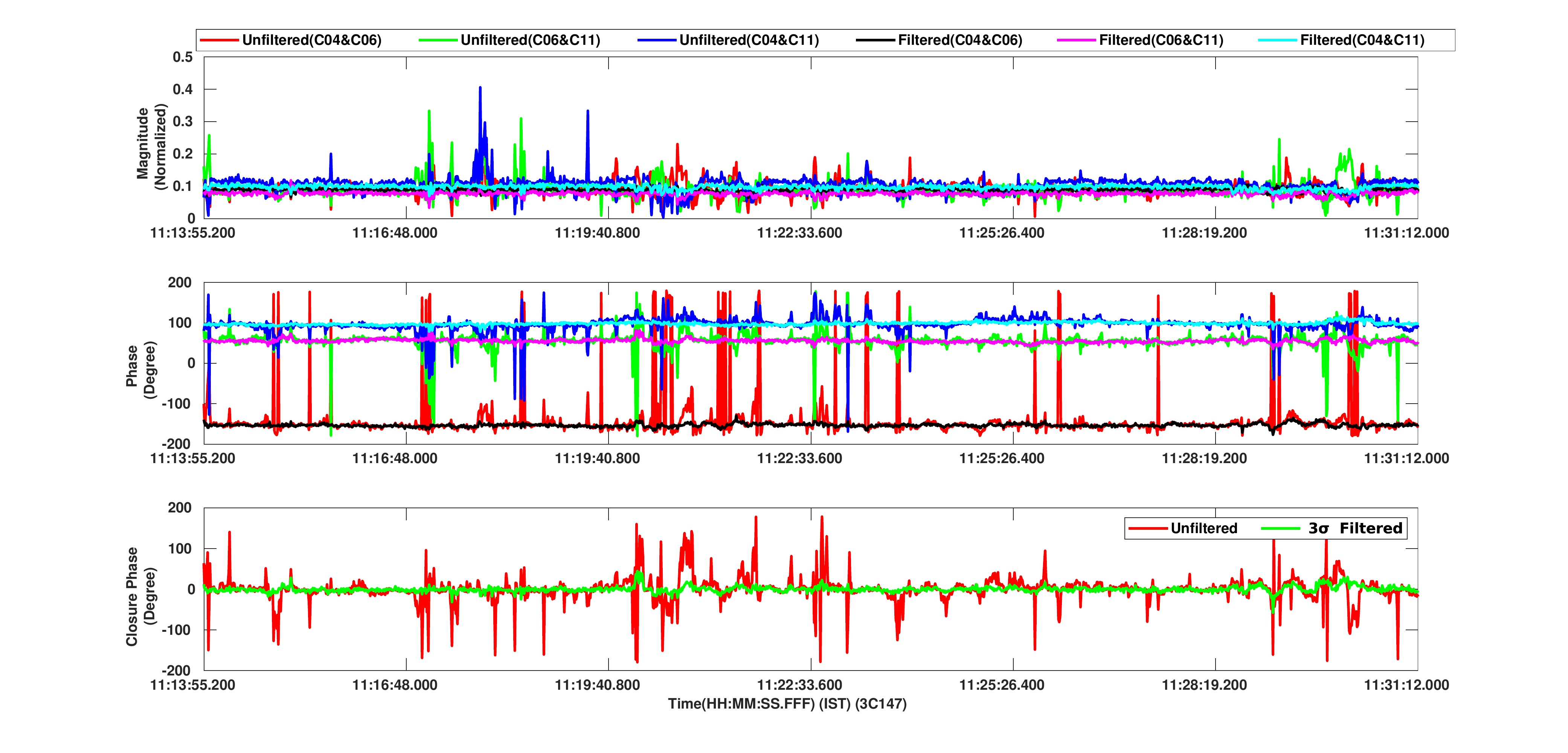}
    \caption{Comparison of closure phase formed by C00, C02, and C04 antennas and their filtered counterparts. The plots shown are for a single spectral channel (corresponding to 431.64 MHz) versus time. The baselines lengths for C00-C02, C02-C04, and C00-C04 corresponding to spectral channel number 700 are 688m, 581m, and 1263m, respectively. The Y-axis of the upper subplot shows the normalized cross-correlation magnitude for the unfiltered and filtered baselines, and the middle subplot shows the cross-correlation phase. The lower subplot shows the closure phase for the unfiltered and filtered antennas.}
    \label{ccf}
\end{figure*}

A comparison of the closure phase for a calibrator source (unresolved) observed in uGMRT Band-3 is shown in Fig. \ref{ccf}. The observation was carried out in 1:2 digital copy mode.

\section{Impact on astronomical data}\label{sec:astro}

\subsection{Continuum imaging}
We expect that removing broadband RFI can lead to better data and, thus improved images. We have presented the effects on the visibilities and images in \citep{ruta2020ncra}\footnote{\url{http://www.ncra.tifr.res.in/~ruta/files/NCRA-tech-report-R1401.pdf}}. Here we provide an example of one such test to illustrate the impact.

\begin{table}[]
    \centering
\tabularfont
\caption{Summary of uGMRT observations.}\label{tab:imagingobs} 
\begin{tabular}{lccccc}
\topline
Date& Band &Time&Source&\\
& & min & &\\\midline
11 Aug. 2019&3&20 &3C48& \\
& &35 & 0116-208&\\
& &180 &A2744 &\\
13 Aug. 2019&4& 30&3C48&\\
& & 45& 0116-208&\\
& & 240& A2744&\\
\hline
\end{tabular}
\end{table}

The facility of recording filtered and unfiltered data simultaneously (1:2 mode) 
was used to test the impact of real-time RFI mitigation on continuum imaging.
Data from half of the GMRT array were recorded, and the same data after filtering were recorded in the 1:2 mode. The observation was carried out like a typical continuum observation involving scans on primary calibrator, secondary calibrator, and target source. 
Since the broadband RFI 
 is correlated at the short baselines, the test included imaging of an extended source. The target Abell 2744 was chosen due to extended emission \citep{2021A&A...654A..41R}  
that would be detected even with half of the array. We carried out the test at bands 3 and 4. A summary of the observations is provided in Table~\ref{tab:imagingobs}. 
 \subsubsection{Data analysis}
 For radio interferometric data analysis, Common Astronomy Software Applications \citep{2007ASPC..376..127M} of the NRAO is the standard package. In order to follow identical analysis steps on the unfiltered and filtered data, we used an automated pipeline. We analysed our data using an early version of CAPTURE (CASA Pipeline-cum-Toolkit for uGMRT Data Reduction)  \citep{2021ExA....51...95K} that allowed exact data analysis steps to be performed on the unfiltered and the filtered data. The standard steps of flagging (removing bad data from non-working antennas and RFI), calibration and imaging were performed. Self-calibration was carried out, as is standard practice when making continuum images, to improve the sensitivity of the image.
 
 \subsubsection{Results}
\begin{figure}
    \centering
    \includegraphics[trim =1.5cm 1.5cm 2.7cm 1cm,clip,width =8 cm]{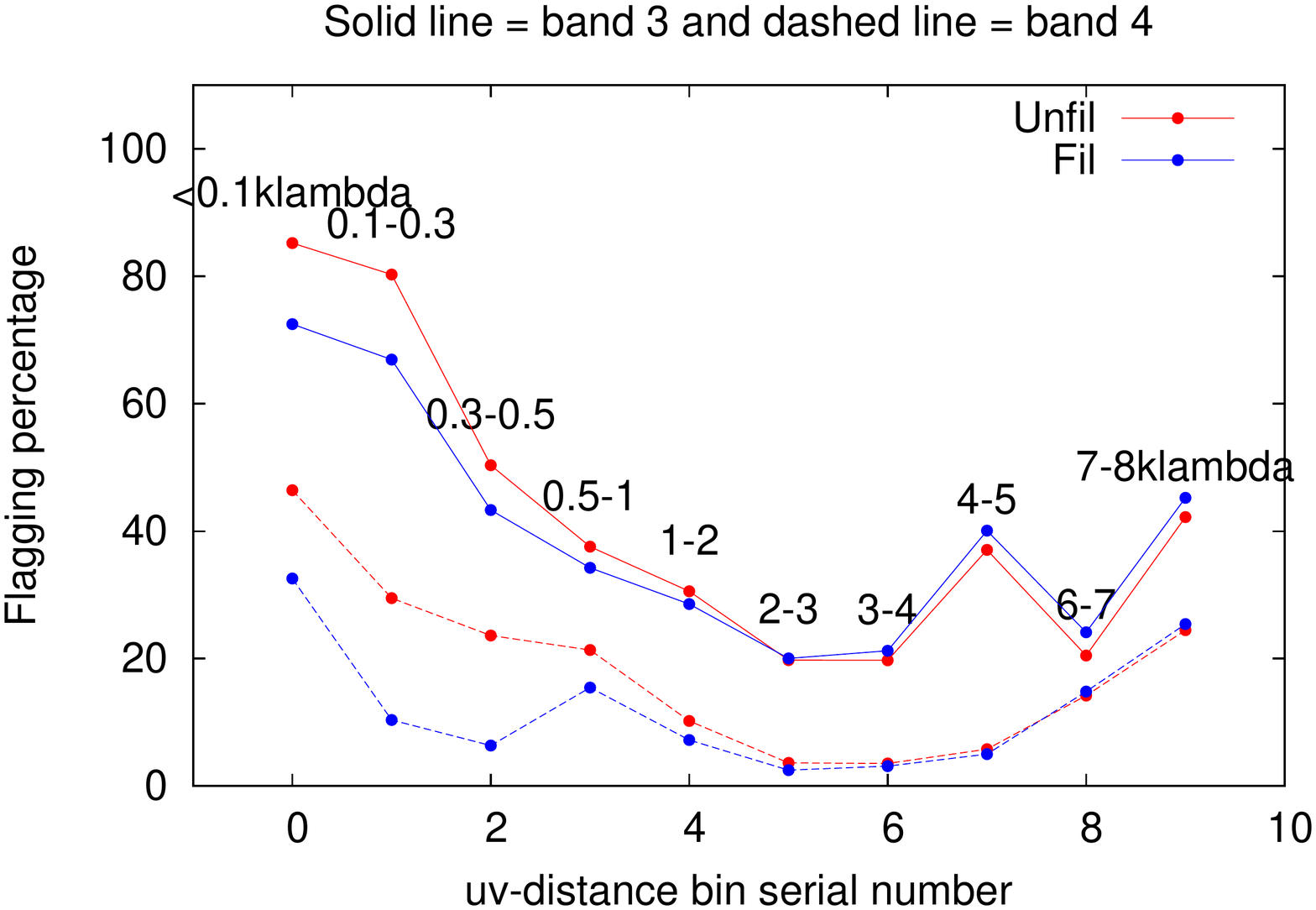}
    \caption{Flagging percentages versus uv-distance bins are plotted. The uv-distance bins in units of kilolambda are labelled in the plot. The solid lines are for band 3, and the dashed ones are for band 4. The unfiltered data are shown in red and the filtered in blue.}
    \label{fig:rfi-percent-comp}
\end{figure}

\begin{figure*}
    \centering
    \includegraphics[width=17cm]{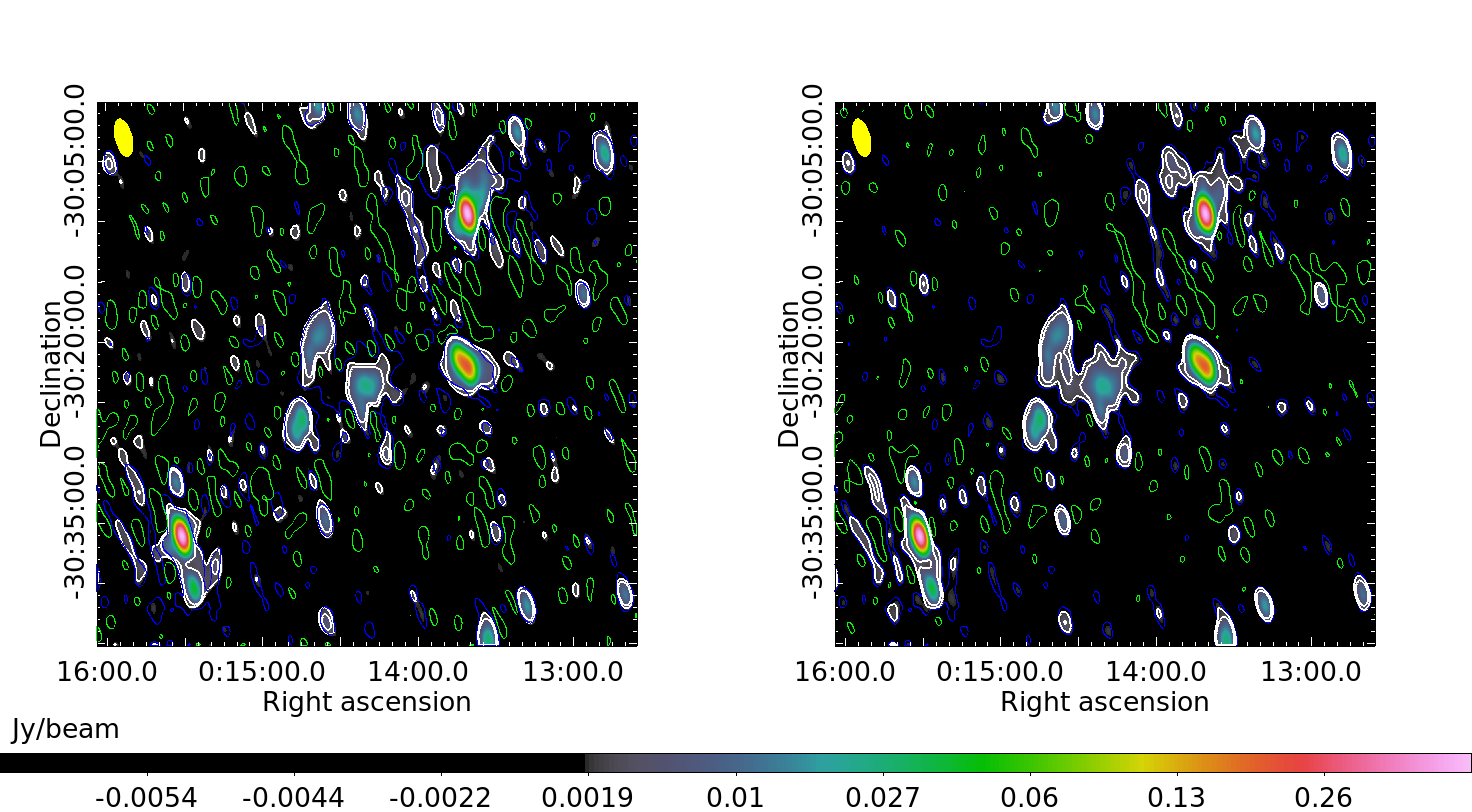}
    \caption{Image of A2744 made with the use of baselines $< 2$klambda at band 3 using unfiltered data (left)
and filtered data (right). The contours are drawn at identical levels in the two panels. The green contours
are at $-0.0014$; blue is at 0.0014 Jy beam$^{-1}$. This represents a level of 2$\sigma$ where $\sigma$ is the RMS in the
filtered image. The white contours are at 0.0021, and 0.0042 Jy beam$^{-1}$ and represent 3 and 6 $\sigma$ levels. The coordinates and the colour scale (Jy beam$^{-1}$) are matched in the two panels. The beam size of $97.8''\times44.3''$, position angle of $13.3^{\circ}$ is shown as a filled ellipse at the top left in the panels.}
    \label{fig:shortuvimgs}
\end{figure*}

 Here we present results that illustrate the impact of real-time filtering on the data and images. We compared the percentage flagging over uv-distance bins in the visibilities at bands 3 and 4. The uv-range was divided into ten bins, and in each bin, the flagging percentage was obtained using the CASA task ``visstat''. The result is presented in 
 Fig.~\ref{fig:rfi-percent-comp}. The flagging percentage is highest in the shortest baselines. For both the bands, the filtered data resulted in $\sim 15\%$ less flagging. We also studied the impact on continuum images at bands 3 and 4. 
Here we present radio images that are made only using short baselines ($<2 $klambda) at band 3 (Fig.~\ref{fig:shortuvimgs}). The RMS (root mean square) values in the images from the unfiltered data is $1.1$ mJy beam$^{-1}$ and that from the filtered data is $0.7$ mJy beam$^{-1}$.
Comparing the images shows that the filtered image can recover the extended central source with a better signal-to-noise ratio. Compared to the RMS noise level in the filtered image, the unfiltered image has artefacts (Fig.~\ref{fig:shortuvimgs}).

\subsection{Spectral-line Observations}

We carried out spectral-line observations to test the effect of real-time RFI filtering. The target source was the HI 21cm absorption line of the z$=2.192$ galaxy towards TXS2039+187 \citep{kanekar2013search}, observed simultaneously with the GMRT Software Backend (GSB; 32 MHz backend at GMRT) \citep{roy2010real} and the GWB, with $\sim$40 minutes of on-source time. The GSB used a bandwidth of 2.067 MHz with 512 channels, while the GWB used a bandwidth of 3.125 MHz and 4096 channels; the bands were centered on the red-shifted 21cm line frequency, 445.0 MHz. Fig.~\ref{spectral_line_plot_overlay} shows an overlay plot from the GSB and GWB observations. The narrow absorption line is clearly seen in both spectra. However, the GWB spectral baseline appears cleaner than the GSB spectral baseline. The RMS noise of the GWB spectrum is $20\%$ better than that of the GSB spectrum after smoothing the spectra to the same spectral resolution.

\begin{figure}
    \centering
       \includegraphics[trim =3.5cm 5cm 3.0cm 5cm,width =6.0cm]{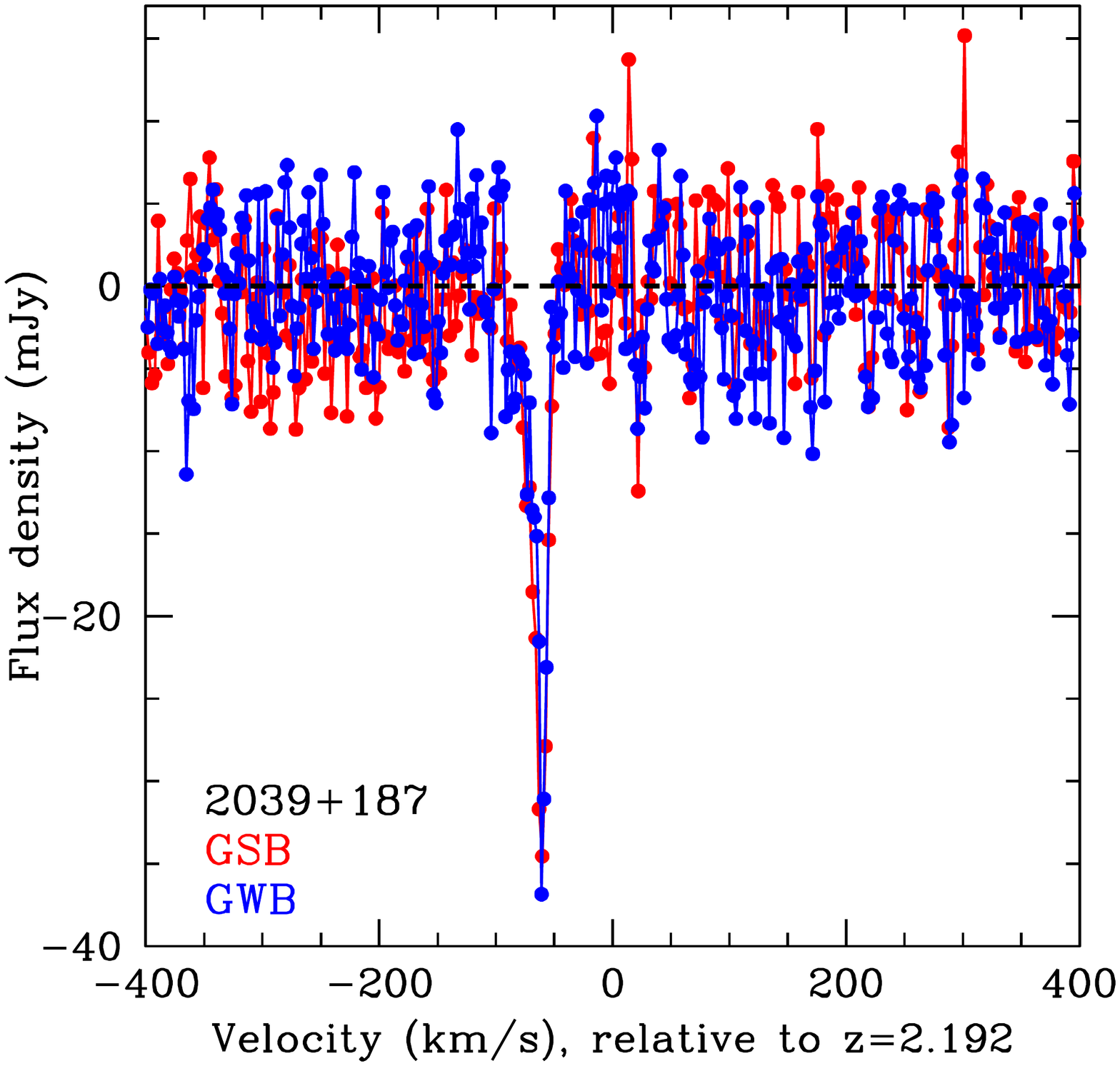}
      \caption{Overlay plot of spectral line observations using the GSB and GWB.}
    \label{fig:spectral_line_overlay}
\end{figure}

\subsection{Time-domain Astronomy}

\begin{figure*}
    \centering
    \includegraphics[trim =1.5cm 0.5cm 1cm 1cm,width =13 cm]{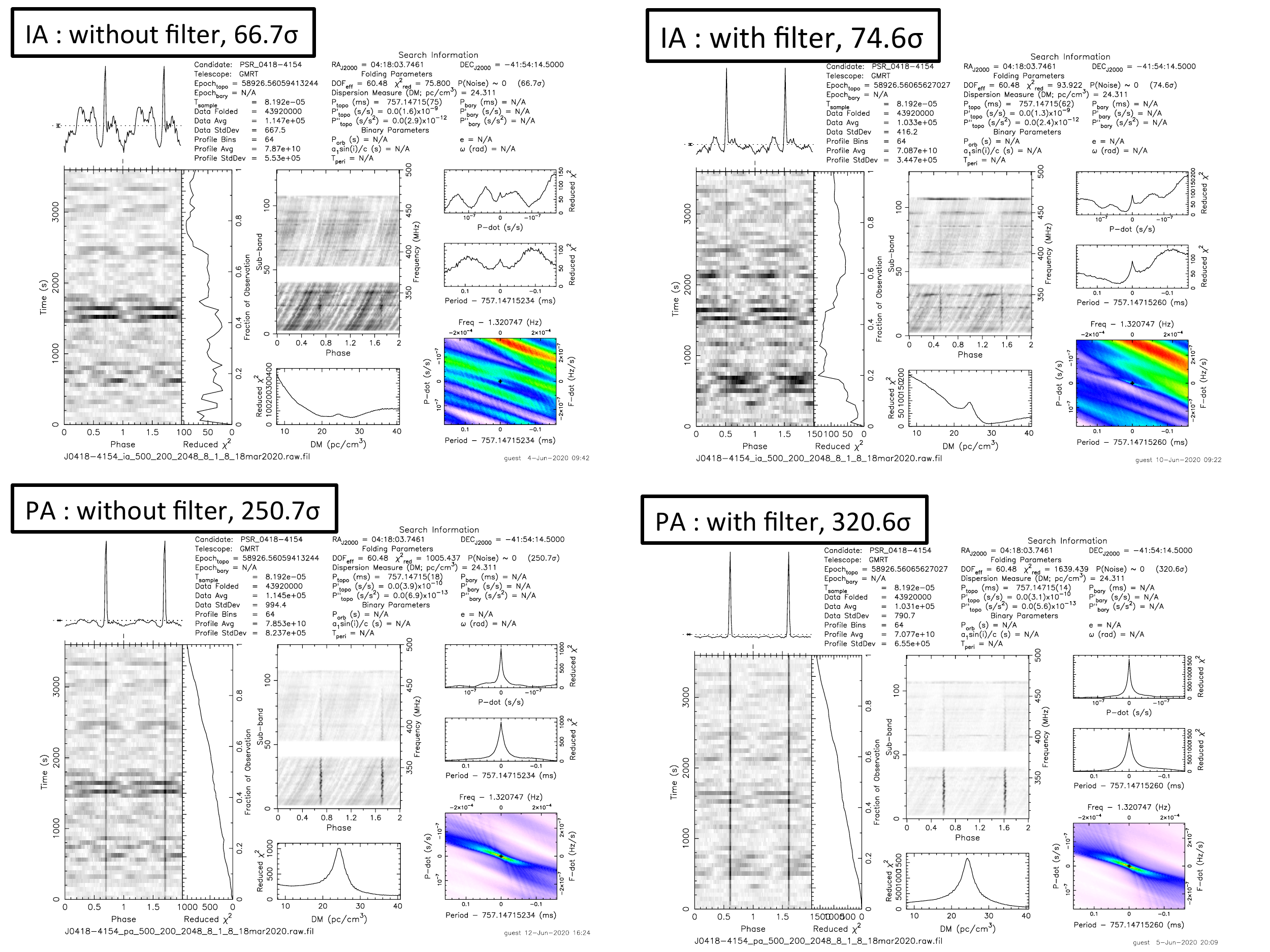}
    \caption{Prepfold (part of standard pulsar analysis software PRESTO \citep{ransom2002fourier}) detection plot for PSR J0418-4154:  a) IA-without filter, b) IA-with filter, c) PA-without filter and d)  PA-with filter. Significant improvement in detection significance while using real-time RFI filtering is observed.}
    \label{fig:pulsar_plot}
\end{figure*}

We conducted tests at multiple epochs to investigate the online RFI filtering system's performance for beamformer observations using the uGMRT. 
We recorded filtered and unfiltered data from the beamformer observations in this experiment. We also recorded both incoherently phased array (IA) and coherently phased array (PA) data. We observed multiple pulsars at uGMRT Band-2 and Band-3 in this setup at multiple observing epochs using the total intensity mode. Thus, we record four simultaneous data files for each target pulsar for a) IA-without filter, b) IA-with filter, c) PA-without filter, and d)  PA-with filter. The detection of Pulsar J0418-4154, a pulsar discovered in the GMRT High-Resolution Southern Sky (GHRSS) survey \citep{bhattacharyya2016gmrt}, for these four scenarios is shown in Fig.~\ref{fig:pulsar_plot}. Significant improvement of SNR is seen with online filters in both IA and PA mode observations. To summarise, we observed improved detection in IA mode with the RFI filtering for the observations at uGMRT band-3. However, we observe comparable or better signal-to-noise ratio (SNR) detection in PA mode with RFI filtering. We note that the observed improvement in the detection significance will depend on the RFI condition of the particular epoch and the threshold chosen for the RFI filtering. We observe marginal or no improvement in the SNR in the case of our band-2 observations, which could be because a significant fraction of data gets flagged for band-2, which was severely affected by RFI.

\begin{figure}
    \centering
    \includegraphics[trim=0cm 0.2cm 0.2cm 1.25cm, clip, width = 8cm]{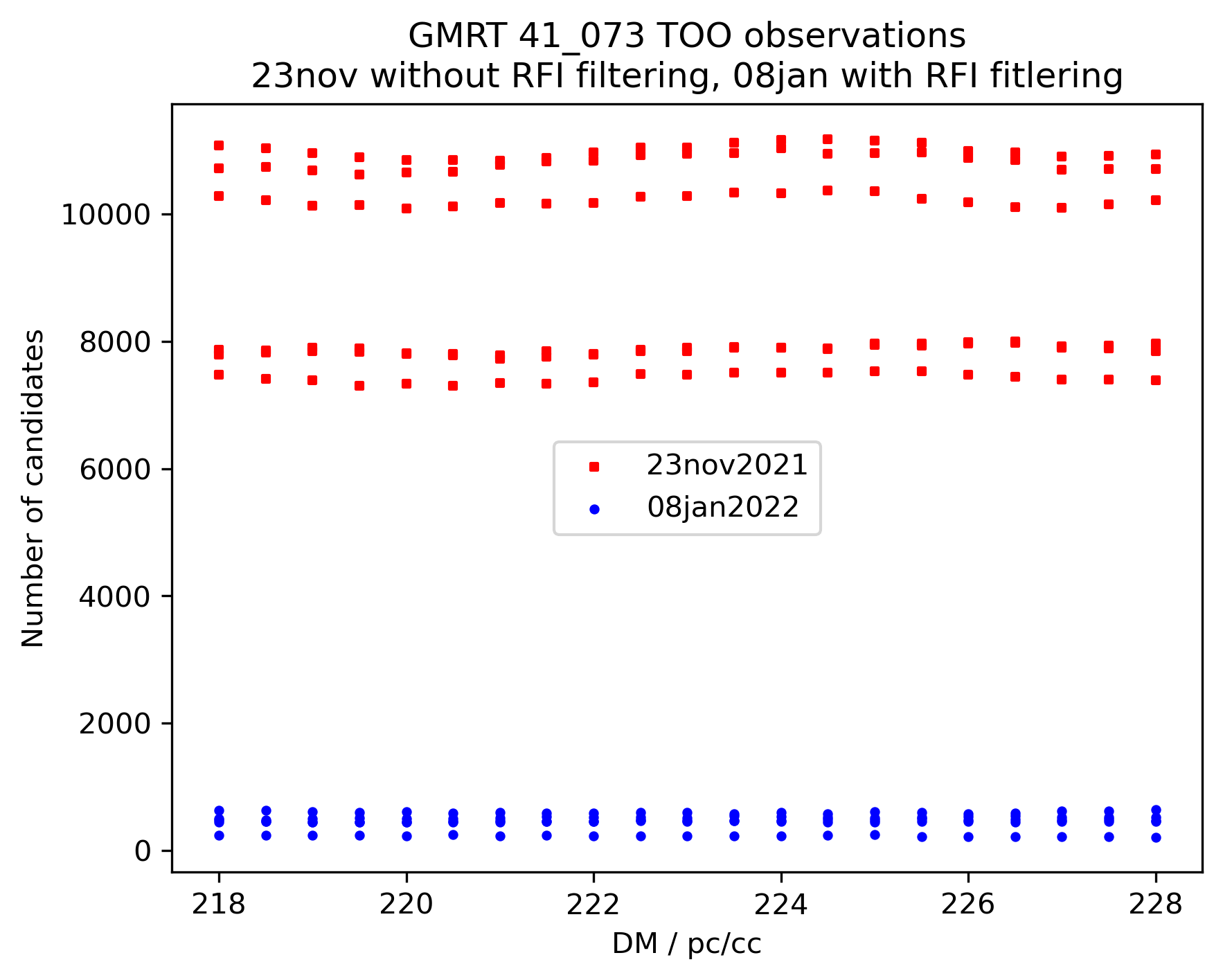}
    \caption{Plot showing FRB candidates in observing runs with (blue circles) and without (red squares) using the RFI filtering.}
    \label{fig:frb}
\end{figure}

The RFI filtering system was also used for Fast Radio Bursts (FRB) observations. Data from a typical FRB observation undergoes de-dispersion, similar to pulsar observations, followed by detection and selection of candidates. RFI produces many spurious candidates, and the RFI filter dramatically reduces the number of candidates. In a particular FRB search observation using uGMRT band-4 and 14 antennas, the RFI filter resulted 20 times fewer candidates (i.e. from 10000 to 500), significantly improving the chances of detecting an actual FRB event in the presence of RFI, which would otherwise spawn mostly spurious candidates (Fig.~\ref{fig:frb}).

\begin{table*}
\begin{center}
\caption{Details of the observing setup for PSR J0418$-$4154}
\vspace{0.3cm}
\label{discovery}
\begin{tabular}{|l|c|c|c|c|c|c|c|c|c|c|c|c|c|c|c|c}
\hline
Pulsar name    & Period  & Dispersion measure & Band  & F$_{res}$$^\dagger$ & T$_{res}$$^\ddagger$ & T$_{obs}$$^\ast$\\
               & (ms)    & (pc~cm$^{-3}$)     & (MHz) & (kHz)               &      ($\mu$s)        &  (hr) \\\hline
PSR J0418$-$4154 & 757.11  & 24.5             & 300$-$500 &   97            & 81                   &   1 hr \\\hline         
\end{tabular}
\end{center}
$^\dagger$: Time resolution\\
$^\ddagger$: Frequency resolution\\
$^\ast$: Observing duration\\

\end{table*}

\section{Current Usage Statistics }\label{sec:usage}

The RFI filtering system is released on a shared-risk basis from the GTAC (GMRT Time Allocation Committee) observing Cycle 38 (April - October 2020). Since then, several observations in uGMRT bands - 2, 3, and 4 have used the RFI filtering system. Observing continuum radio sources and detection of weak pulsars and FRBs have primarily used RFI filtering. We present band-wise statistics for the GTAC cycle 40 in Table  \ref{gtacusage}.

The filtering usage has increased in the last two cycles, reaching almost $50 - 80\%$ in bands 2 and 3. A recent publication \cite{schellenberger2022unusually} cited the RFI filtering system. The filter was also used in the GHRSS survey \footnote{\url{http://www.ncra.tifr.res.in/~bhaswati/GHRSS.html}}.

RFI filtering system will be released as a regular feature for bands 3 and 4 from April 2022 (GTAC Cycle 42), and band-2 will continue on shared risk. New features based on the uGMRT user community feedback are being introduced to the filtering system while improving our understanding of the effects on astronomical data.

The percentage of RFI detected and filtered for a given threshold depends on the number of outliers in the data. For a pure noise signal, about 0.03\% of data are filtered for a 3$\sigma$ threshold. Table \ref{perflag} provides the typical average percentage RFI filtered for uGMRT bands at 3 and 4 sigma thresholds. These percentage values are likely to change in the case of a very strong RFI and are dependent on the RFI environment. We recommend the filtering threshold based on the tradeoff from the benefits of filtering to the artefacts introduced due to excessive filtering.

User-programmable features of the RFI filtering system are the selection of threshold factor ($n$ value in the threshold equation shown in Sec.~\ref{sec:techreview}), replacement options, bypass mode, and applying filtering to either or both the polarization. The replacement options include replacement by digital noise, constant value, and threshold value. The recommended settings of threshold and replacement are provided in the GTAC User Document 

The typical flagging percentages from various tests and GTAC observations are 4-5\% for band-2, 1-2\% for band-3, 0.5-1.5\% for band-4, and $<0.5$\% for band-5.

\begin{table*}[]
\centering
\caption{Comparison of average percentage flagging for different uGMRT bands and noise at 3 and 4 $\sigma$ thresholds}\label{perflag}
\vspace*{0.2cm}
 \begin{tabular}{cccc}
 \hline
 \hline
   Signal Type & 3$\sigma$ & 4$\sigma$\\
  \hline
    Noise  & 0.03 &0.001\\
  2  & 4-5 & 1.5-3\\  
  3 & 2-3 & 1-1.5 \\
  4 & 0.5-1.5 & $<$ 1 \\
  5 & $<$ 0.5 & $<$ 0.1 \\
   \hline
 \end{tabular}
\end{table*}

 \begin{table*}[]
\centering
\caption{Real-time RFI filtering usage during GTAC observations in Cycle 40 (April-October 2021) for uGMRT Band-2 (120-240 MHz) and Band-3 (250-500 MHz). The table shows the total approved GTAC duration and RFI filtering usage (in hours and percentage of total duration) along with the type of observation.}\label{gtacusage}
\vspace*{0.2cm}
 \begin{tabular}{cccc}
 \hline
 \hline
   Band & Duration (Hrs.) & Filtering (Hrs.)($\%$) & Obs. Type\\
  \hline
    2   & 30 &24 ($80\%$)  & Continuum ($100\%$)\\
  3    & 418 &203 ($49\%$) & Continuum ($55\%$), Pulsar($45\%$)\\  
   \hline
 \end{tabular}
\end{table*}

The facility of recording the flagging counter, as mentioned in Sec.~\ref{sec:techreview}, can be used during the observations. The commands for recording the RFI flagging count can be added to the command file used for GTAC observations. Command file generation using online tool \footnote{\url{http://www.ncra.tifr.res.in/~secr-ops/cmd/cmd.html}}, if the online RFI filtering option is activated, adds the commands every 5 min. interval in the scan period.

\begin{figure*}
    \centering
    \includegraphics[width=13cm]{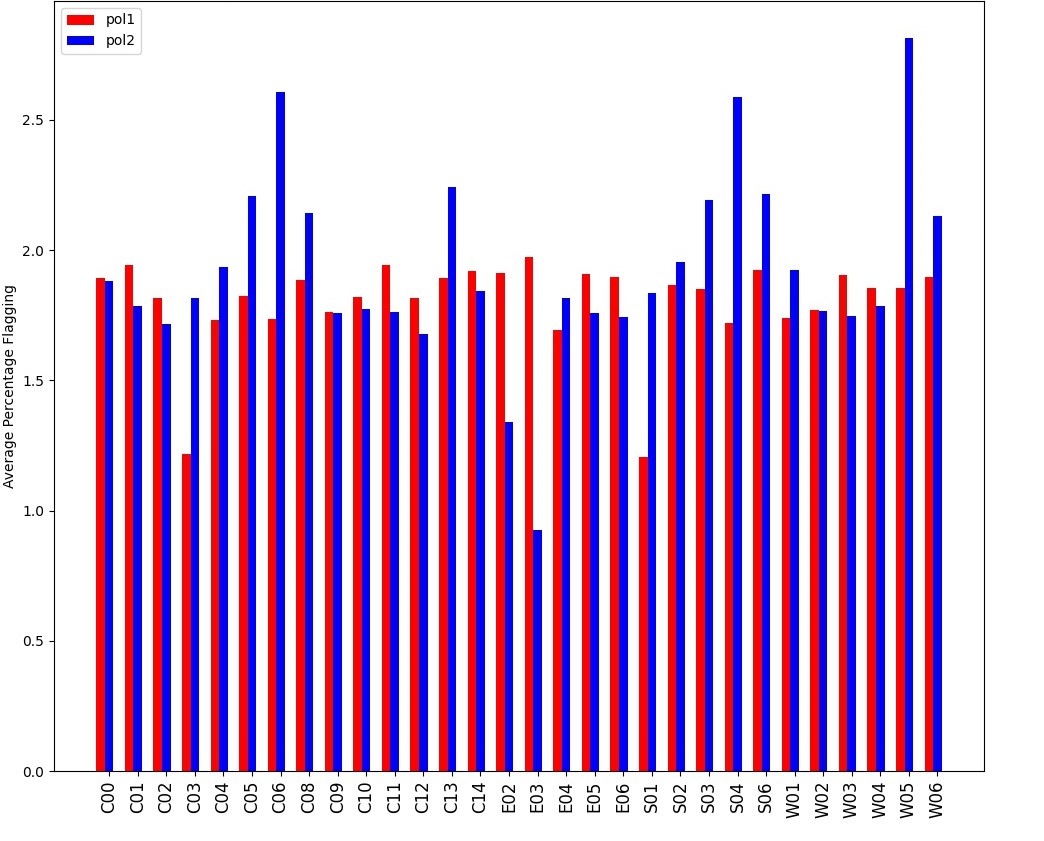}
    \caption{Antenna-wise average percentage of RFI flagging for an observing run. }
    \label{rficntr}
\end{figure*}

The recorded RFI flagging data can be converted to percentage flagging for each counter epoch. We have developed a utility \footnote{\url{http://www.gmrt.ncra.tifr.res.in/~kdbuch/counter_page/counter_flagging.html}} to plot the percentage flagging of individual counter epochs, the average for all the antennas, and a single antenna for multiple epochs. An example of percentage flagging for GMRT antennas is shown in Fig \ref{rficntr}.

\section{Relevance to the SKA}\label{sec:skarel}

SKA is expected to receive RFI from various terrestrial and extra-terrestrial sources (like communication satellites) for SKA-Low and SKA-Mid telescopes even though the location is remote with low population density and radio-quiet zones \citep{tingay2020survey, otto2016characterising} characterising. The RFI environment at the SKA sites is mainly dominated by narrowband RFI though there is a possibility that in future, there may be broadband RFI at the SKA antennas that are far from the core (SKA-TEL-SKO-0000492\footnote{\url{https://www.skatelescope.org/wp-content/uploads/2018/12/SKA-TEL-SKO-0000492-01-RFICharacterSKA1SignalChainDesignConsider-part-1-signed.pdf}}).
The report also deals with signal chain design considerations and has proposed using threshold-based detection of RFI. The report also mentions the need to excise strong RFI in the pre-correlation domain and generate a flag from the detector. Detection of RFI in high-resolution data through antenna-level buffering is also proposed in \citet{nelles2019applications} for the SKA-Low.

The current implementation of the broadband RFI filtering at the uGMRT can  be extended for usage with other observatories such as the SKA.

As mentioned in Section \ref{sec:techreview}, some trade-offs were considered while designing and testing the RFI filtering. These were specific to the FPGA and the hardware board used in the GWB. The most computationally challenging aspect is the real-time implementation of the median calculation, which was carried out using the histogram method \citep{buch2019real}. An alternative to the sorting required for computing the median can be achieved through the histogram method, optimal for integer arithmetic. The complexity of this technique primarily depends on the input bit-width and window size \citep{buch2019real}. For example, for a q-bit input bit-width, the complexity would scale as $2^q$. The histogram method is suitable for input bit-width less than or equal to 8 and a window size of up to 65,536 samples.

The technique operates on raw voltage and is an antenna-based detection and filtering scheme making it convenient to scale the computational and resource requirements to the SKA level. The design is not directly dependent on the receiver's bandwidth; the only constraint is the real-time operation. The system architecture is modular, and the filtering parameters are user-configurable. 

The SKA-1 signal processing system is planned to be implemented on an FPGA platform \citep{2020SPIE11445E..89V}. The following are the possible trade-offs for implementing the real-time system on hardware or software platforms planned to be used for the SKA.

\begin{itemize}
\item \textbf{Computing standard deviation on discrete intervals:} In this option, the standard deviation is computed over a particular window and applied to the following several windows. The computation is carried out at discrete intervals. This trade-off would provide relaxation in the real-time constraint. The gap between the standard deviation computation can be decided on the variability in the RFI environment and on the real-time constraints provided by the digital signal processing system.

\item \textbf{Filtering-only mode:} Here, the robust standard deviation can be calculated externally or at the start of the observation and supplied to the filtering system. The filtering operates on real-time data. This scheme would ease the resource and timing constraints on the filtering system. The interval at which the standard deviation can be applied depends on the variability in the RFI environment and other system-related and external parameters affecting the input power levels.

\item \textbf{Computing standard deviation in time-multiplexed mode:} In the case of multiple antennas on a single computing platform, the standard deviation can be computed in a time-multiplexed fashion, i.e. while the computation is going on for one antenna, the other antennas hold the previous value of standard deviation. Filtering happens in real time for all the antennas. This scheme helps meet resource requirements as there is one standard deviation computation block for every $N$ antenna. Only one standard deviation computation block is required for $N$ antennas that are processed on a single FPGA or a GPU node.
\end{itemize}

\subsection{Real-time RFI monitoring}

The RFI filtering system can work in a piggyback mode with only the threshold-based detection to get the RFI statistics per antenna during each observation. The filtering operation can be bypassed if required. The RFI counter can still operate and provide the flagging statistics. The statistics obtained can be used for long-term monitoring of the RFI environment and understanding the levels of RFI in different bands, RFI direction, and the effect of other parameters on varying levels of interference. Depending on the system's location, it can be used to monitor broadband or narrowband RFI.
\subsection{Extension to mitigate narrowband RFI}
The concept of real-time RFI filtering developed for uGMRT can be extended for mitigating narrowband signals in the spectral domain and on integrated visibilities or beam data. The algorithm is generic with user-defined parameters and can be used in other telescopes, particularly the SKA. The previous sections showed that the system could filter impulsive RFI at the highest time resolution and improve the quality of astronomical results with minimal data loss. This section provides recommendations on the usage of the filtering system for the SKA.

Depending on the type of RFI and the Interference-to-Noise Ratio (INR) \citep{series2013techniques}, parameters like the estimation interval or the detection domain (voltage or power) need to be modified to achieve optimum performance. Although the system at uGMRT has been tested and released for broadband RFI, a similar algorithm by \citet{buch2016towards} has been applied to offline time-frequency data in the post-correlation domain for mitigation of narrowband RFI. The algorithm can also be applied to beamformer data for mitigating broadband and narrowband RFI for time-domain astronomy observations \citep{buch2016towards}. Thus, the real-time technique can be used on raw and processed signals at different signal-processing systems. As an example, a possible approach for implementing the technique in the spectral domain on the CPU platform is shown in \citet{buch2016real}.

\section{Summary and conclusions}\label{sec:conclusions}
RFI mitigation is essential for all radio band observatories due to the unprecedented sensitivities that are being achieved and the rising interference levels. We reviewed the real-time RFI filtering implemented at the uGMRT. The technique, implementation, testing methods and impact on continuum, spectral-line and time-domain data were described. The release of this system for the users of the uGMRT and the growing usage were presented. The relevance of such schemes in the context of the SKA was also presented. A summary of these sections is provided below:
\begin{itemize}
    \item The real-time RFI filtering technique is implemented on FPGAs and uses a MAD-based scheme for identifying impulsive RFI. It allows replacing the samples identified as RFI with digital noise, a constant or zero. The user can choose the threshold for RFI identification and the replacement option.
    \item The testing of the real-time RFI filtering on the uGMRT has been possible using the specially developed modes of recording filtered and unfiltered data simultaneously. 
    \item The engineering tests showed that suppression of RFI improved SNR, cross-correlation function, and closure phase relations.
    \item The astronomical data in continuum mode show a dramatic improvement in the short baseline data and can increase the fidelity of images of extended sources. Spectral line data show an improved estimation of the noise in line-free channels.
    \item In the time domain astronomy, real-time RFI filtering has been tested for observing pulsars and FRBs and used in major projects like the GHRSS survey.
    \item The real-time RFI filtering at the uGMRT has been released regularly for bands 3, 4, and 5 since April 2022. During the last observing cycle (April- October 2021), $\sim$ $50\%$ observations in band-3 and $80\%$ observed in band-2 used the real-time RFI filtering system.
    \item There is an overlap in the observing frequencies of uGMRT and SKA Low. The mitigation scheme developed for uGMRT is similar to that proposed for RFI mitigation in the SKA signal processing chain.
    \item uGMRT RFI mitigation system is modular and amenable for implementation on hardware and software platforms. Various tradeoffs were provided to take care of real-time computational constraints and operate in piggyback mode.
\end{itemize}
In light of growing RFI levels, mitigation methods that will result in minimal data loss will be necessary. Implementing the real-time RFI filtering scheme at the uGMRT will serve as a pathfinder of schemes for the SKA. 

\section*{Acknowledgements}

We thank the past team members who worked on the real-time RFI filtering system for uGMRT. We acknowledge the GMRT control room, Operations group, and Backend group members for their help in carrying out the RFI filtering test observations. We are thankful to Pravin Raybole and the RFI group at GMRT for discussions on potential sources of RFI and their properties. We thank Bhaswati Bhattacharyya and Jayanta Roy for their help in analyzing the pulsar test results, Nissim Kanekar for the spectral-line test results, and Visweshwar Ram Marthi for the details on FRB-related results provided in this paper. We would also like to thank NCRA faculty members for their suggestions and feedback on the filtering system.
\vspace{-1em}

\bibliographystyle{apj}
\bibliography{rev-jaasample}


\end{document}